\documentclass[a4paper]{article}

\usepackage{INTERSPEECH2019}
\usepackage{siunitx}
\title{Who said that?: Audio-visual speaker diarisation of real-world meetings}
\name{Joon Son Chung, Bong-Jin Lee, Icksang Han}
\address{Naver Corporation, South Korea}
\email{\{joonson.chung,bongjin.lee,icksang.han\}@navercorp.com}

\begin{document}

\maketitle

\begin{abstract}

The goal of this work is to determine ‘who spoke when' in real-world meetings. The method takes surround-view video and single or multi-channel audio as inputs, and generates robust diarisation outputs. 

To achieve this, we propose a novel iterative approach that first enrolls speaker models using audio-visual correspondence, then uses the enrolled models together with the visual information to determine the active speaker. 

We show strong quantitative and qualitative performance on a dataset of real-world meetings. The method is also evaluated on the public AMI meeting corpus, on which we demonstrate results that exceed all comparable methods. We also show that beamforming can be used together with the video to further improve the performance when multi-channel audio is available.

\end{abstract}

\section{Introduction}

Over the recent years, there has been a growing demand to be able to record and search human communications in a machine readable format. There has been significant advances in automatic speech recognition due to the availability of large-scale datasets~\cite{panayotov2015librispeech,barker2018fifth} and the accessibility of deep learning frameworks~\cite{Abadi16,paszke2017automatic,Vedaldi15}, but to give the transcript more  meaning beyond just a sequence of words, the information on ‘who spoke when' is crucial. 

Speaker diarisation, the task of breaking up multi-speaker audio into single speaker segments, has been an active field of study over the years. Speaker diarisastion can mostly be addressed as a single-modality problem where only the audio is used, but there are also a number of papers that have used additional modalities such as video. Previous works on speaker diarisation, both audio and audio-visual, can be divided into two strands.

The first is based on speaker modelling (SM) which uses the assumption that each individual has different voice characteristics. Traditionally, speaker models are constructed with Gaussian mixture models (GMMs) and i-vectors~\cite{dehak2011front,cumani2013probabilistic,matvejka2011full}, but more recently deep learning has been proven effective for speaker modelling~\cite{variani2014deep,lei2014novel,ghalehjegh2015deep,snyder2017deep,snyder2018x}.
In many systems, the models are often pre-trained for the target speakers~\cite{hung2008towards,biagetti2016robust} and are not applicable to unknown participants.
Other algorithms are capable of adapting to unseen speakers by using generic models and clustering~\cite{friedland2012icsi,sell2018diarization}. 
There are also a number of works in the audio-visual domain that are based on feature clustering~\cite{friedland2009multi,sarafianos2016audio}.

The second strand uses a technique referred to as sound source localisation (SSL), which is claimed to demonstrate better performance compared to the SM-based approaches according to a recent study~\cite{rozgic2010multimodal}, particularly with powerful beamforming methods such as SRP-PHAT~\cite{dibiase2000high}. However, SSL-based methods are only effective when the location of speakers are either fixed or known. Therefore SSL has been used as parts of audio-visual methods, where the location of the identities can be tracked using  the visual  information~\cite{schmalenstroeer2009fusing}. This approach is dependent on the ability to effectively track the participants. A recent paper~\cite{cabanas2018multimodal} combines SSL with a visual analysis module that measures motion and lip movements, which is relevant to our work.

A number of works have combined SM and SSL approaches using independent models for each type of observation, then fused these information with a probabilistic framework based on the Viterbi algorithm~\cite{schmalenstroeer2009fusing} or with Bayesian filtering~\cite{rozgic2010multimodal}. 

In this paper, we present an audio-visual speaker diarisation system based on self-enrollment of speaker models that is able to handle movements and occlusions. 
We first use a state-of-the-art deep audio-visual synchronisation network to detect speaking segments from each participant when the mouth motion is clearly visible. This information is used to enroll speaker models for each participant, which can then determine who is speaking even when the speaker is occluded. By generating speaker models for each participant, we are able to reformulate the task from an unsupervised clustering problem into a supervised classification problem, where the probability of a speech segment belonging to every participant can be estimated.
In contrast to the previous works that compute likelihoods for each type of observation before the multi-modal fusion, the audio-visual synchronisation is used in the self-enrollment process. 
Finally, when multi-channel microphone is available, beamforming is employed to estimate the location of the sound source, then the spatial cues from both modalities are used to further improve the system's performance. 
The effectiveness of the method is demonstrated on the internal dataset of real-world meetings and the public AMI corpus. 

This paper is organised as follows. In Section~\ref{subsec:baseline}, we first describe the audio-only baseline system  based on state-of-the-art methods for speech enhancement, activity detection and speaker diarisation. Section~\ref{subsec:multimodal} introduces the proposed audio-visual system. Finally, Section~\ref{sec:exp} describes the datasets and the experiments in which we demonstrate the effectiveness of our method on the public AMI dataset.

\section{System description}
\label{sec:system}

\subsection{Audio-only baseline} 
\label{subsec:baseline}

The baseline system provided for the second DIHARD challenge is used as our audio-only baseline.
The system takes key components from the top-scoring systems in the first DIHARD challenge and shows state-of-the-art performance on audio-only diarisation.

\subsubsection{Speech enhancement} The speech enhancement is based on the system used by USTC-iFLYTEK in their submission to the first DIHARD challenge~\cite{sun2018speaker}. The system uses Long short-term memory (LSTM) based speech denoising model trained on simulated training data. It has demonstrated significant improvements in deep learning-based single-channel speech enhancement over the state-of-the-art, and the authors have shown its effectiveness for diarisation with a second-place result in the first DIHARD challenge.

\subsubsection{Speech activity detection} 

The speech activity detection baseline uses WebRTC~\cite{johnston2012webrtc} operating on enhanced audio processed by the speech enhancement baseline. 

\subsubsection{Speaker embeddings and diarisation} 
\label{subsec:baseline_dia}
The diarisation system is based on the JHU Sys4 used in their winning entry to DIHARD I, with the exception that it omits the Variational-Bayes refinement step.
Speech is segmented into 1.5 second windows with 0.75 second hops, 24 MFCCs are extracted every 10ms, and a 256-dimensional x-vector is extracted for each segment. The extracted vectors are scored with PLDA (trained with segments labelled only for one speaker) and clustered with AHC (average score combination at merges). 

The x-vector extractor and PLDA parameters were trained on the VoxCeleb~\cite{Nagrani17} and VoxCeleb2~\cite{Chung18a} datasets with data augmentation (additive noise), while the whitening transformation was learned from the DIHARD I development set~\cite{sell2018diarization}. We use the pre-trained model released by the organisers of the DIHARD challenge.

The system is not designed to handle overlapped speech, and additional speakers are counted as missed speech in evaluation.

\subsection{Multi-modal diarisation} 
\label{subsec:multimodal}

\begin{figure}[!b]
\centering 
\includegraphics[width=1\linewidth]{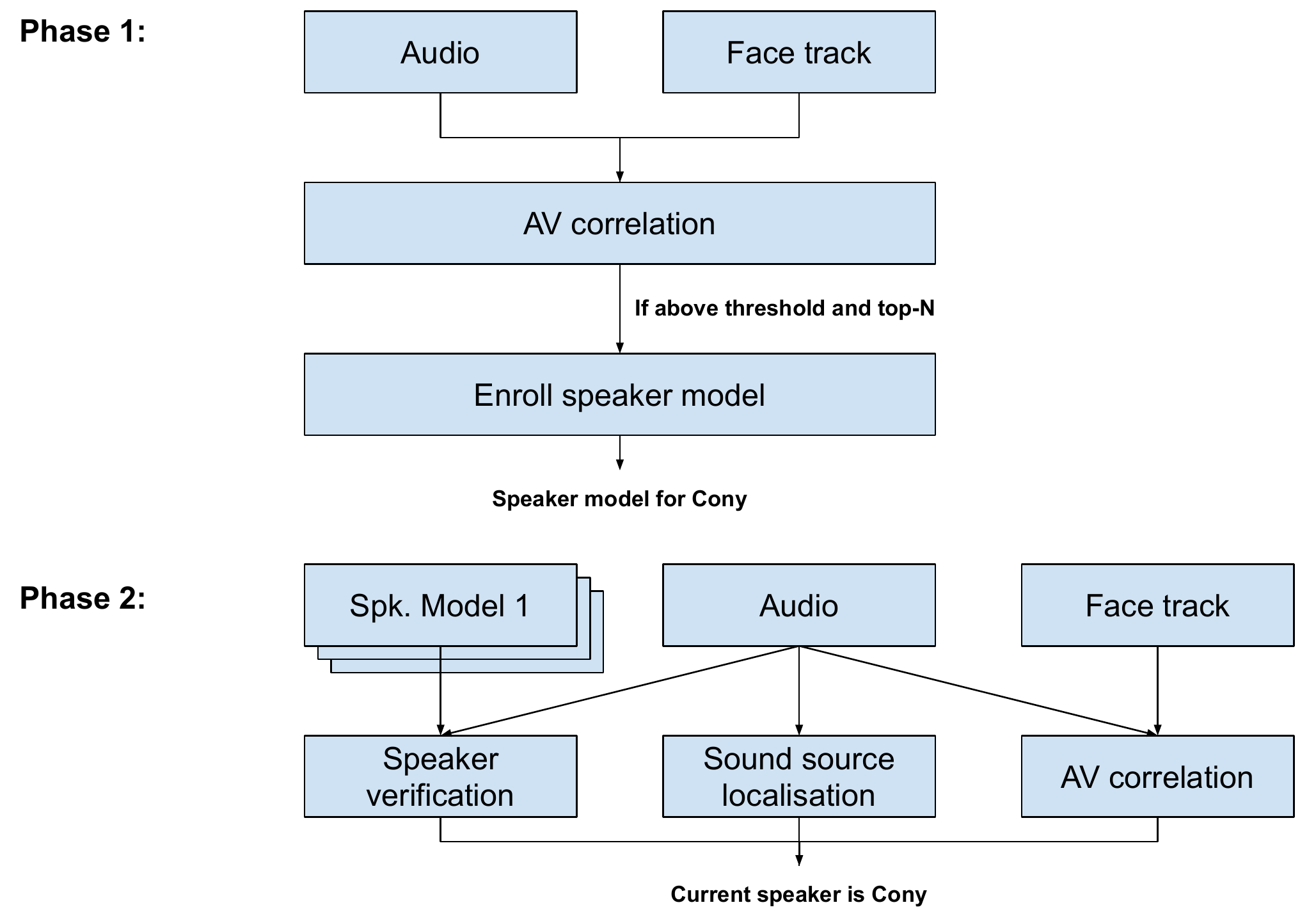}
\caption{Pipeline overview.}
\label{fig:pipeline} 
\end{figure}

The audio processing part of the audio-visual system shares most of the baseline methods described above: the speech enhancement and speech activity detection modules are identical to that in the baseline system, and for experiments on the AMI corpus, we also use the pre-trained x-vector model used by the JHU system to extract speaker embeddings.

Three modes of information are used to determine the current speaker in the video. The pipeline is summarised in Figure~\ref{fig:pipeline}  and described in the following paragraphs.

\subsubsection{Audio-to-video correlation}
\label{subsec:avcorr}
Cross-modal embeddings of the audio and the mouth motion are used to represent the respective signals. The strategy to train this joint embedding is described in~\cite{chung2018perfect}, but we give a brief overview here. 

The network consists of two streams: the audio stream that encodes Mel-frequency cepstral coefficients (MFCC) inputs into 512-dimensional vectors; and the video stream that encodes cropped face images also into 512-dimensional vectors. The network is trained as a multi-way matching task between one video clip and $N$ audio clips. Euclidean distances between the audio and video features are computed, resulting $N$ distances. The network is trained with a cross-entropy loss on the inverse of this distance after passing through a softmax layer, so that the similarity between matching pairs is greater than non-matching pairs. 

The cosine distance between the two embeddings is used to measure correspondence between the two inputs. Therefore, we expect small distance between the features if the face image corresponds to the current speaker and in-sync and large distance otherwise. Since the video is from a single continuous source, we assume that the AV offset is fixed throughout the session. The embedding distance is smoothed over time using a median filter in order to eliminate outliers.

\subsubsection{Speaker verification}
\label{subsec:sr}

We develop speaker models for each individual (identified in Sec.~\ref{subsec:facerec}) so that the active speaker can be determined even when audio-visual synchronisation cannot be established due to occlusion.

The audio-visual pipeline (Sec.~\ref{subsec:avcorr}) is run over the whole video in advance, in order to determine $N$ most confident speaking segments (each of 1.5 seconds) for each identity. In our case we use $N$=10, and if there are fewer than $N$ confident segments above a AV correlation threshold, we only use the segments whose correlation is above the threshold. These are used to enroll the speaker models.

For the experiments on the AMI dataset, we use the x-vector network (described in Sec.~\ref{subsec:baseline_dia}) to extract speaker embeddings, so that the results can be compared like-for-like to the baseline.

For the experiments on the internal meeting dataset, we use a deeper ResNet-50 model~\cite{He16} also trained on the same data as the baseline. The deeper model is used here since its features generalise better to this more challenging dataset compared to the shallower x-vector model.

At test time, speaker embeddings are extracted by computing features over 1.5-second window, moving 0.75 seconds at a time, in line with the baseline system. By comparing the embeddings at each timestep to the enrolled speaker models, the likelihood of the speech segment belonging to any individual can be estimated. Even without any visual information at inference time, this now becomes a supervised classification problem, which is typically more robust compared to unsupervised clustering.

\subsubsection{Sound source localisation}
\label{subsec:amb}

Besides the speaker embeddings, the direction of the sound source can provide useful cues on who is speaking.

Recordings from the 4-channel microphone from the GoPro camera can be converted to Ambisonics B-Format using the GoPro Fusion Studio software. By solving the B-format representations for azimuth $\theta$ and elevation $\phi$, the direction of the audio source can be estimated for each audio sample. The direction for every video frame  is determined by generating a histogram of all $\theta$ values over a $\pm0.5$  second period with bin size of $\ang{10}$.

For the AMI videos, the Time Delay of Arrival (TDOA) information is calculated using the BeamformIt~\cite{anguera2007acoustic} package. As with the internal dataset, the direction of arrival is also computed with a histogram of $\theta$ values over a $\pm0.5$  second period. However, only 4 bins of $\ang{90}$ is used since the video is split over 4 cameras and the exact geometry between them is unknown.

The likelihood of the audio belonging to any person at a given time correlates to the angle between the estimated audio source and the face detection in the video for the identity in question.

\subsection{Multi-modal fusion} 
\label{subsec:fusion}

Each of the three modalities (AV correlation, speaker models, direction of audio) give confidence scores for each speaker and timestep. These scores are combined into a single confidence score for every speaker and timestep using a simple weighted fusion as stated below, where $C_{sm}$ is the confidence score from the speaker model, $C_{avc}$ is the score from the AV correspondence and $\phi$, $\theta$ are the directions of the face and the estimated DoA of audio, respectively. When the identity is not visible on the camera, the second and third terms are put to zero.

\begin{equation}
C_{overall} = C_{sm} + \alpha * C_{avc} + \beta * cosine(\phi - \theta)
\end{equation}

\subsection{Implementation details}

\subsubsection{Face detection and tracking}
\label{subsec:facetrack}
A CNN face detector based on Single Shot MultiBox Detector (SSD)~\cite{Liu16} is used to detect face appearances on every frame of the video. This detector allows faces to be tracked across wide range of poses and lighting conditions. A position-based face tracker is used to group individual face detections into face tracks.

\subsubsection{Face recognition}
\label{subsec:facerec}
The method requires face images for each participant so that they can be identified and tracked regardless of their position in the room. This can be from user input or from their profile images. The face images for all participants are supplied to the VGGFace2~\cite{cao2017vggface2} network, and their embeddings are stored. For each face track detected (Sec.~\ref{subsec:facetrack}), face embeddings are extracted using the VGGFace2 network and compared to each of the $N$ stored embeddings, so that they can be classified into one of $N$ identities. We apply the constraint that co-occurring face tracks at any point in time cannot be of the same identity.

\begin{table*}[!ht] 
\vspace{-15pt}
\footnotesize
\begin{center}
\begin{tabular}{ |l|l|l|  |r|r|r|r| |r|r|r|r| } 
 \hline

Method & Dataset & Input & \multicolumn{4}{c||}{System VAD} & \multicolumn{4}{c|}{Reference VAD}    \\  \hline
    \multicolumn{3}{|c||}{Measure}        & MS & FA & SPKE & DER  & MS & FA & SPKE & DER  \\ \hline                                        
JHU Baseline~\cite{sell2018diarization} & {\bf ES All}  & 1ch       &   10.5 & 6.6  & 12.8  & 30.0             			& 5.6  & 0.0  & 12.2 & 17.8 \\ \hline                                        
{\bf Ours (SM)} & {\bf ES All}   & 1ch+V      &   10.5 & 6.6  & 6.7  & 23.8             				& 5.6  & 0.0  & 7.9 & 13.5 \\ \hline                                        
{\bf Ours (SM+AVC)} & {\bf ES All}    & 1ch+V      &   10.5 & 6.6  & 4.0 & 21.1              & 5.6  & 0.0  & 4.8 & 10.4 \\ \hline 
{\bf Ours (SM+AVC+SSL)} & {\bf ES All}    & 8ch+V      &   10.5 & 6.6  & {\bf 2.8} & {\bf 19.9}              & 5.6  & 0.0  & {\bf 3.6} & {\bf 9.2} \\ \hline \hline

Cabanas et al.~\cite{cabanas2018multimodal} & {\bf ES WB}    & 8ch+V      &   - & -  & - & 27.2              		& - & - & - & -   \\ \hline 
{\bf Ours (SM+AVC)} & {\bf ES WB}    & 1ch+V      &   11.4 & 7.1  & 4.9 & 23.3              	& 6.1 & 0.0 & 5.9 & 12.0 \\ \hline 
{\bf Ours (SM+AVC+SSL)} & {\bf ES WB}    & 8ch+V      &   11.4 & 7.1  &  {\bf 3.8} & {\bf 22.3}              	& 6.1 & 0.0 & {\bf 4.9} & {\bf 10.9} \\ \hline \hline

Cabanas et al.~\cite{cabanas2018multimodal} & {\bf ES NWB}    & 8ch+V      &   - & -  & - & 20.6              	& - & - & - & -   \\ \hline 
{\bf Ours (SM+AVC)} & {\bf ES NWB}    & 1ch+V      &   9.5 & 5.7  & 2.7 & 17.8              & 5.1 & 0.0 & 3.3 & 8.4  \\ \hline
{\bf Ours (SM+AVC+SSL)} & {\bf ES NWB}    & 8ch+V      &   9.5 & 5.7  & {\bf 1.4} & {\bf 16.6}              & 5.1 & 0.0 & {\bf 1.9} & {\bf 7.0}  \\ \hline \hline

JHU Baseline~\cite{sell2018diarization} & {\bf IS All}     & 1ch     &  11.2 & 4.0  & 10.2 & 25.4 						& 6.5 & 0.0 & 11.2 & 17.7     \\ \hline                                        
{\bf Ours (SM)} & {\bf IS All}     & 1ch+V     &   11.2 & 4.0  & 7.6 & 22.9 													& 6.5 & 0.0 & 8.8 & 15.3     \\ \hline                                        
{\bf Ours (SM+AVC)} & {\bf IS All}     & 1ch+V     &   11.2 & 4.0  & 6.2 &  21.3								& 6.5 & 0.0 & 7.1 & 13.6     \\ \hline 
{\bf Ours (SM+AVC+SSL)} & {\bf IS All}     & 8ch+V     &   11.2 & 4.0  & {\bf 4.9} & {\bf 20.0} 										& 6.5 & 0.0 & {\bf 5.8} & {\bf 12.3}     \\ \hline \hline

Cabanas et al.~\cite{cabanas2018multimodal} & {\bf IS WB}    & 8ch+V      &   - & -  & - & 32.3 						& 	- & - & - & -   \\ \hline 
{\bf Ours (SM+AVC)} & {\bf IS WB}    & 1ch+V      &   13.3 & 5.1  & 7.7 & 26.1 									&7.9 & 0.0 & 8.9 & 16.9 \\ \hline 
{\bf Ours (SM+AVC+SSL)} & {\bf IS WB}    & 8ch+V      &   13.3 & 5.1  & {\bf 6.5} & {\bf 24.8} 									&7.9 & 0.0 & {\bf 7.8} &  {\bf 15.7} \\ \hline  \hline

Cabanas et al.~\cite{cabanas2018multimodal} & {\bf IS NWB}    & 8ch+V      &   - & -  & - & 21.7              	& - & - & - & -   \\ \hline 
{\bf Ours (SM+AVC)} & {\bf IS NWB}    & 1ch+V      &   9.3 & 2.8  & 4.8 &  16.8							&  5.3 & 0.0 & 5.4 & 10.6  \\ \hline 
{\bf Ours (SM+AVC+SSL)} & {\bf IS NWB}    & 8ch+V      &   9.3 & 2.8  & {\bf 3.4} & {\bf 15.5} 							&  5.3 & 0.0 & {\bf 4.0} & {\bf 9.3}   \\ \hline \hline

JHU Baseline~\cite{sell2018diarization} & {\bf Internal}     & 1ch     &  1.8 & 4.5  & 72.2 &78.6						& 0.0 & 0.0 & 73.3 & 73.3      \\ \hline                                        
{\bf Ours (SM)} & {\bf Internal}     & 1ch+V     &   1.8 & 4.5  & 24.8 & 31.1 													& 0.0 & 0.0 & 25.6 & 25.6     \\ \hline                                        
{\bf Ours (SM+AVC)} & {\bf Internal}      & 1ch+V     &  1.8 & 4.5  & 18.7 &  25.0								& 0.0 & 0.0 & 19.4 & 19.4     \\ \hline 
{\bf Ours (SM+AVC+SSL)} & {\bf Internal}      & 8ch+V     &   1.8 & 4.5  & {\bf 13.1} & {\bf 19.4} 										& 0.0 & 0.0 & {\bf 13.7} & {\bf 13.7}     \\ \hline

\end{tabular}                                               
\end{center}
\normalsize
\caption{Diarisation results (lower is better). The results are on the AMI dataset except for the last four rows. {\bf WB}: Whiteboard;  {\bf NWB}: No whiteboard; {\bf $X$ch+V}: $X$ channel audio + video; {\bf SM}: Speaker Modelling; {\bf AVC}: Audio Visual Correspondence; {\bf SSL}: Sound Source Localisation; {\bf MS}: Missed Speech; {\bf FA}: False Alarm; {\bf SPKE}: Speaker Error; {\bf DER}: Diarisation Error Rate.}
\label{tab:ami_results}
\vspace{-5pt}
\end{table*}

\section{Experiments}
\label{sec:exp}

The proposed method is evaluated on two independent datasets: our internal dataset of meetings recorded with 360$^{\circ}$ camera, and the publicly available AMI meeting corpus. Each will be described in the following paragraphs.
 
\subsection{Internal meeting dataset}
\label{subsec:internaldata}

The internal meeting dataset consists of audio-visual recording of regular meetings in which no particular instructions are given to the participants with regard to the recording of the video. The meetings form parts of daily discussions from the workspace of the authors and are not set up in any way with the diarisation task in mind. A large proportion of the dataset consists of very short utterances with frequent speaker changes, providing an extremely challenging condition.

\begin{figure}[!t]
\centering 
\includegraphics[width=1\linewidth]{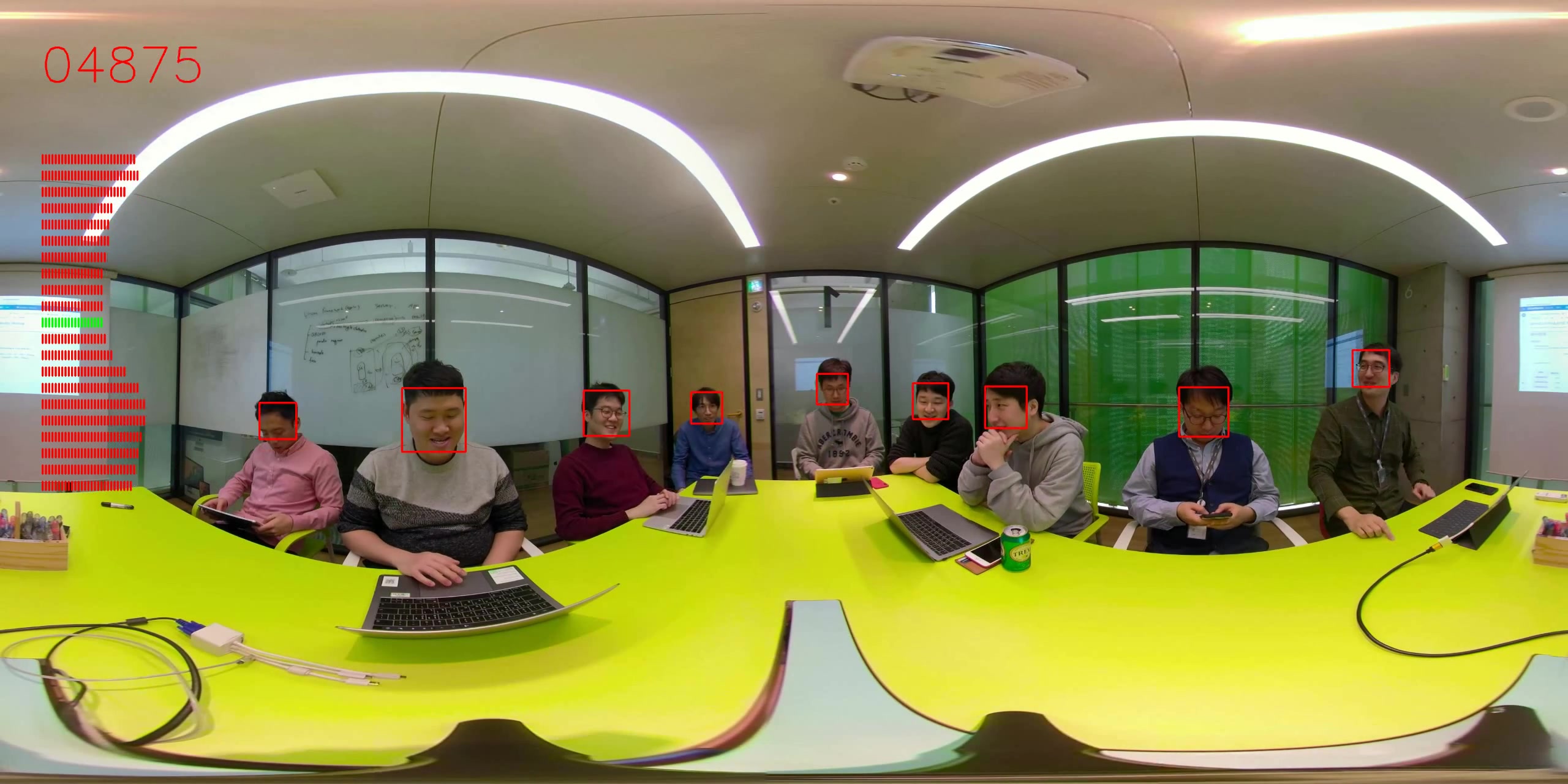}
\caption{Still image from the internal meeting dataset. }
\label{fig:internaldata} 
\end{figure}

The video is recorded using a GoPro Fusion camera, which captures $\ang{360}$ videos of the meeting with two fish-eye lenses. The videos are stitched together into a single surround-view video of 5228x2624 resolution at 25 frames per second. The audio is recorded using a 4-channel microphone at 48 kHz. A still image from the dataset is shown in Figure~\ref{fig:internaldata}.

The dataset contains approximately 3 hours of validation set and 40 minutes of carefully annotated test set. The test video contains 9 speakers. In the case of overlapped speech, we only annotated the ID of main (loudest) speaker. The embedding extractor and the AV synchronisation network are trained on external datasets, and the validation set is only used for tuning the AHC threshold in the baseline system  and the fusion weights in the proposed system.

\begin{figure}[t]
\centering 
\includegraphics[width=0.49\linewidth]{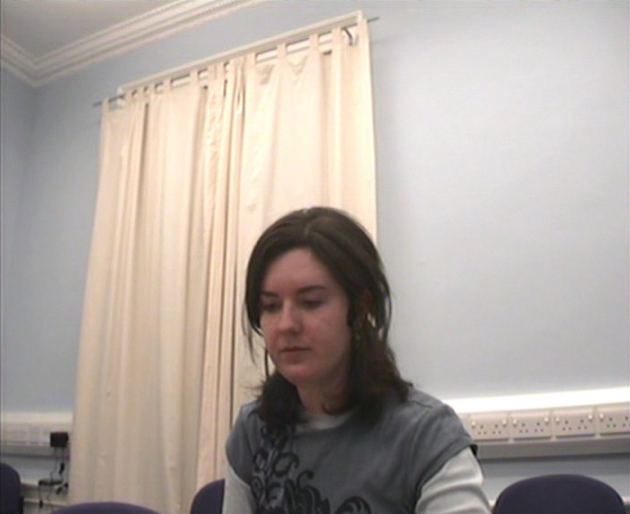}
\includegraphics[width=0.49\linewidth]{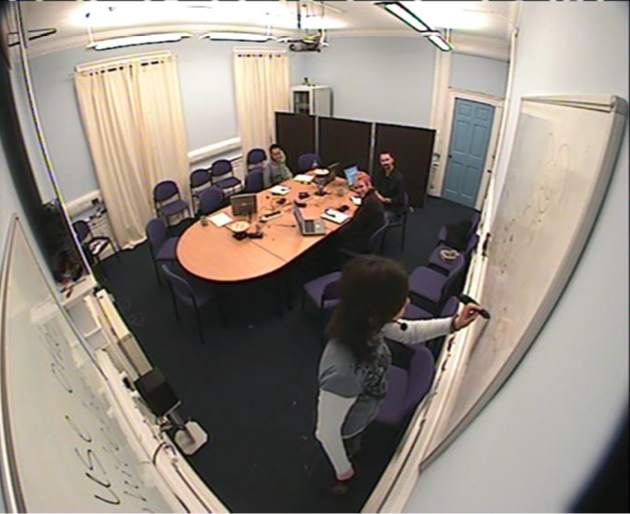}
\caption{Still images from the AMI corpus. }
\label{fig:ami_room} 
\end{figure}

\subsection{AMI corpus}
\label{subsec:ami}
The AMI corpus consists of 100 hours of video recorded across a number of locations and has been used by many previous works on audio-only and audio-visual diarisation. Of the 100 hours of video, we evaluate on meetings in ES (Edinburgh) and IS (Idiap) categories, which contain approximately 30 and 17 hours of video respectively. On the IS videos, 
{\tt IS1002a}, {\tt IS1003b}, {\tt IS1005d}, {\tt IS1007d} were not used in the experiments due to partially missing data.
The image quality is relatively low, with the video resolution of 288x352 pixels.

The audio is recorded from an 8-element circular equi-spaced microphone array with a diameter of 20cm. However, we only use one microphone from the array in most of our experiments. The video is recorded with 4 cameras providing close-up views of each of the meeting's participants, and unlike the internal dataset (Sec.~\ref{subsec:internaldata}), the images are not stitched together. 
%The schematic representation of the setup is shown in Figure~\ref{fig:ami_room}.  

The ES videos is used as the validation set for tuning the thresholds.

\subsection{Evaluation metric}

We use Diarisation Error Rate (DER) as our performance metric. The DER can be decomposed into three components: missed speech (MS, speaker in reference, but not in hypothesis), false alarm (FA, speaker in hypothesis, but not in reference) and speaker error (SPKE, speaker ID is assigned to the wrong speaker). 

The tool used for evaluating the system is the one developed for the RT Diarization evaluations by NIST~\cite{istrate2005nist}, and  includes acceptance margin of 250 ms to compensate for human errors in reference annotation.

\subsection{Results}

Results on the AMI corpus~\cite{carletta2005ami} are given in Table~\ref{tab:ami_results}.
The numbers for meetings where the whiteboard is used are provided separately, so that the results can be compared to~\cite{sell2018diarization}.

Missed speech and false alarm rates are the same across different models for each dataset since we use the same VAD system in all of our experiments. Therefore the speaker error rate (SPKE) is the only metric affected by the diarisation system.

Our speaker model only system (SM) uses the visual information only to find out when to enroll the speaker models, and during inference only uses the audio. Since the audio processing pipeline and the embedding extractor are common across our system and the JHU-based baseline, the performance gain arises  from changing a clustering problem into a classification problem. This alone results in 48\% and 26\% relative improvement in speaker error on the ES and IS  sets, respectively. 

It is also clear from the results that the addition of the AV correspondence (AVC) and sound source localisation (SSL) at inference time both provide boost to the performance. The contributions of these modalities to overall relative performance are 20-40\% and 19-39\% respectively depending on the test set.

%Overlapped speech is not detected in all of the methods and this is accounted for as missed speech (MS).

Note that our results exceed the recent audio-visual method of~\cite{cabanas2018multimodal} across all test conditions by a significant margin, whilst using the same input modalities.  \cite{friedland2010dialocalization} also reports competitive results on a subset of the IS videos (SPKE of 7.3\%, DER of 19.5\% using 4 cameras and 8 microphones), however the results cannot be compared directly to our work since some of the test videos are no longer available at the time of writing this paper.

The speaker error rates are markedly worse on the internal meeting dataset, presumably due to the more challenging nature of the dataset and the larger number of speakers. From the results in Table~\ref{tab:ami_results}, it can be seen that the baseline system does not generalise to this dataset, but the proposed multi-modal systems perform relatively well on this ‘in the wild' data.

\section{Conclusion}

In this paper, we have introduced a multi-modal system which takes advantage of audio-visual correspondence to enroll speaker models. We have shown that speaker modelling with audio-visual enrollment have significant advantages over clustering methods typically used for diarisation. Areas for further research include learnable methods for multimodal fusion, improvements to the speech activity detection (SAD) modules and the combination of audio-visual diarisation and audio-visual speech separation for meeting transcription and for handling overlapped speech.

\vspace{10pt}
\noindent\textbf{Acknowledgment.}
We would like to thank Chiheon Ham, Han-Gyu Kim, Jaesung Huh, Minjae Lee, Minsub Yim, Soyeon Choe and Soonik Kim
for helpful comments and discussion.

\clearpage
\bibliographystyle{IEEEtran}

\bibliography{}

\end{document}